\documentclass[preprint, review,12pt]{iopart}
\usepackage{iopams}
\usepackage[utf8]{inputenc}

\usepackage{graphicx}
\usepackage{hyperref}
\usepackage[square,numbers,sort&compress]{natbib}
\expandafter\let\csname equation*\endcsname\relax
\expandafter\let\csname endequation*\endcsname\relax

\usepackage{commath}

\usepackage[varg]{txfonts}
\usepackage{siunitx}
\usepackage{units}

\usepackage[usenames]{xcolor}
\usepackage[version=3]{mhchem}

\begin{document}
\title{Streamer propagation in humid air}

\author{Alejandro Malagón-Romero and Alejandro Luque}
\address{IAA-CSIC, Granada, Spain}
\ead{amaro@iaa.es}
\vspace{10pt}

\begin{abstract}

We investigate the effect of humidity on the propagation of streamers in air. We present a minimal set of chemical reactions that takes into account the presence of water in a nonthermal air plasma and considers ionization, attachment, detachment, recombination and ion conversion including water cluster formation. We find differences in streamer propagation between dry and humid air that we attribute mostly to an enhanced effective attachment rate in humid air, leading to higher breakdown electric field and threshold field for propagation.  This higher effective attachment rate in humid conditions leads to a faster decay of the conductivity in the streamer channel, which hinders the accumulation of charge in the streamer head.  In some cases a propagating streamer solution still exists at the expense of a smaller radius and lower velocity.  In other cases a high humidity leads to the stagnation of the streamer. We finally discuss how all these statements may affect streamer branching and the dimensions and lifetime of a streamer corona.
\end{abstract}

\section{Introduction}
\label{Introduction}
Streamers are filamentary electric discharges that propagate due to a high electric field at their tip. This field accelerates electrons up to energies high enough to ionize the air in front of the streamer, releasing other electrons and paving the path for its propagation. Streamers constitute the building blocks of other electric discharges, such as the corona sheath around the leader in a long spark discharge, both in laboratory conditions \cite{LesRenardiers1977/Elektra,LesRenardiers1978/Elektra} and in nature (e.g., lightning \cite{Rakov2003/book}). Recent observations suggest that extensive streamer systems, spanning hundreds of meters and containing above \num{e8} streamers are responsible for the powerful radio emissions known as Narrow Bipolar Events (NBEs) \cite{Rison2016/NatCo,Li2022/GeoRL} as well as for transient blue flashes on thundercloud tops \cite{Soler2020/JGRA,Li2021/JGRD}.  Right above the clouds, blue jets also show a streamer phase at the final stages of their propagation, when the air heating is not enough for them to sustain the propagation of a highly conductive hot core (the leader \cite{Wescott1995/GeoRL,Gordillo-Vazquez2009/PSST,Raizer2010/JGRA,Popov2016/JASTP}). Streamer discharges are also present at higher altitudes, forming large-scale electric discharges known as sprites \cite{Franz1990/Sci, Pasko1996/GeoRL, Pasko1998/GeoRL, Raizer1999/ITPS, Cummer2006/GeoRL, Stanley1999/GeoRL, Stenbaek-Nielsen2008/JPhD, Malagon-Romero2020/GRL} that span tens of kilometers in the vertical direction and hundreds of meters or more horizontally.

Many of the discharges mentioned above propagate in humid air, sometimes saturated or even supersaturated.  Therefore, it makes sense to study streamers under such conditions. In 2018, Kostinskiy \textit{et al.} \cite{Kostinskiy2018/temp} carried out experiments where they studied the abrupt elongation of long spark discharges, i.e., a sudden elongation in length preceded and followed by streamer corona bursts. They found indications of positive leader stepping under high humidity conditions. They also pointed out possible signatures of space stems in positive leaders. In 2019 Malagón-Romero and Luque \cite{Malagon-Romero2019/GeoRL} showed the spontaneous formation of space stem precursors in the streamer corona ahead of negative leaders due to an attachment instability. They also stressed the role of water as an electron detachment suppressor favoring the formation of long-lasting plasma non-homogeneous structures.

Water is known to slow down or even supress streamer propagation \cite{Gallimberti1979/JPhys, LesRenardiers1977/Elektra}. Several mechanisms contribute to this behavior: (a) enhancement of the three-body electron attachment
\begin{equation}
\ce{e + O_2 + $M$ -> e + O_2^{-} + $M$},
\label{3body}
\end{equation}
whose reaction rate for $M=\ce{H_2O}$ is 6-10 times greater than that for $M=\ce{O_2}$ \cite{Liu2017/JPhD,Aleksandrov1998/TPhyL}. We neglect the case $M=\ce{N2}$ since the reaction constant is approximately two orders of magnitude smaller than for $M=\ce{O_2}$, see e.g. \cite{Kossyi1992/PSST}.  (b) Electron detachment inhibition due the formation of negative water clusters of the kind \ce{O_2^-(H_2O)_$n$}, where usually $n=1$--3 are the most relevant cases \cite{Gallimberti1979/JPhys} and (c) enhancement of the electron-ion recombination rate through the formation of positive water clusters \ce{O_2^+(H_2O)_$n$} with $n=$1--3 \cite{Aleksandrov1998/TPhyL}.

Numerical simulations of single streamer discharges have been used to study the inception of pulsed discharges \cite{Sun2014/JPhD}, its propagation {\cite{Dhali1985/PhRvA,Kulikovsky1995/JPhD,Pancheshnyi2008/JCoPh,Luque2012/JCoPh} as well as associated processes such as electron runaway and the emission of X-rays \cite{Chanrion2010/JGRA/c,Kohn2017/GeoRL, Ihaddadene2015/GeoRL,Lehtinen2018/JGRD} or the formation of luminous and isolated patterns in the streamer wake \cite{Malagon-Romero2019/GeoRL}. However, most of these studies are performed in pure nitrogen or dry air conditions.

  Here we study the propagation under non-uniform electric fields of single streamer discharges of both polarities in air with several humidity conditions. Then, we describe the effects on characteristic streamer magnitudes such as its velocity, radius and peak electric field. We focus on the conditions (humidity, background electron density, applied electric field) that allow a streamer to emerge from a needle-like initial seed. As we discuss below, the increased effective attachment rate increases the minimum electric field required for the inception of a streamer. Our results also describe a relatively simple chemical model that efficiently incorporates the main effects that water vapour imposes on non-thermal electrical discharges. We conclude by speculating about the implications for the propagation of long spark discharges.

\section{Streamer discharge model}
\label{sec:model}

In this work we simulate the propagation of positive and negative streamers in dry and humid air under laboratory conditions of \SI{300}{\kelvin} and \SI{1}{atm}. Dry air is modeled as a mixture of $79\%$ of \ce{N_2} and $21\%$ \ce{O_2} in volume. In our simulations under humid conditions we proportionally decrease the content fraction of nitrogen and oxygen in the mixture to give room to a water volume content of $1.5\%$ and $3 \%$ (saturated air at \SI{300}{\kelvin}) corresponding to, respectively, \SI{12}{g / m^{3}} and \SI{24}{g / m^{3}}. Our chemical scheme is based on the work by Luque \textit{et al.} \cite{Luque2017/PSST}, with some additions listed in Table \ref{tab:reactions}.

\begin{table}[h]
	\label{tab:reactions}
	\begin{centering}
		\begin{tabular}{lclcc}
			& Reaction & & Rate coefficient ($\unit{m^{3} / s}$ or $\unit{m^{6} / s}$) & Reference\tabularnewline
			\hline
			\ce{e + H2O & -> & 2e + H2O+} & $f\left(x_{0}\right)$ & \textit{Bolsig+}\tabularnewline
			\ce{e + H2O & -> & H- + OH} & $f\left(x_{0}\right)$ & \textit{Bolsig+}\tabularnewline
			\ce{e + H2O & -> & OH- + H} & $f\left(x_{0}\right)$ & \textit{Bolsig+}\tabularnewline
			\ce{e + H2O + O2 & -> & O2- + H2O} & $f\left(x_{0}\right)$ & \textit{Bolsig+}\tabularnewline
			\ce{H- + O2 & -> & e + HO2} & $1.2\cdot10^{-15}$ & \cite{Komuro2013/JPhD}\tabularnewline
			\ce{H- + H2O & -> & OH- + H2}& $3.8\cdot10^{-15}$ & \cite{Komuro2013/JPhD}\tabularnewline
			\ce{O- + H2O & -> & OH- + OH} & $6\cdot10^{-19}$ & \cite{Fehsenfeld1974/JCP}\tabularnewline
			\ce{H- + O2 & -> & O- + OH} & $10^{-17}$  & \cite[p.~185]{McEwan1975/Book} \tabularnewline
			\ce{H- + O2 & -> & O2- + H}  & $10^{-17}$ & \cite[p.~185]{McEwan1975/Book} \tabularnewline
		\end{tabular}
		\par\end{centering}
	\caption{List of the reactions added to the chemical scheme in \cite{Luque2017/PSST} }
\end{table}

The electron and ion densities evolve according to the drift-diffusion-reaction (\textit{ddr} equations:
\begin{equation}
	\label{eq:fluid}
	\frac{\partial n_s}{\partial t} + \nabla\cdot\left(\pm n_s \mu_s \mathbf{E} - D_s\nabla n_s \right) = S_{ph} + C_s,
\end{equation}
where $n_s$, $\mu_s$, $D_s$ and $C_s$ respectively denote the density, mobility, diffusion coefficient and chemical source term for the species $s$. The different sign preceding the drift term accounts for the electric charge of the species $s$. $S_{ph}$ denotes the photoionization term that has been derived from \cite{Zhelezniak1982/HTemS} and was computed using a Helmholtz approach \cite{Luque2007/ApPhL}  with Bourdon’s three-term parameters \cite{Bourdon2007/PSST}. This term only applies to electrons and \ce{O_2^+}. Mobility, diffusion and reaction rate coefficients are computed with \textit{Bolsig+} and depend on the local electric field \cite{Hagelaar2005/PSST}. The cross sections feeding \textit{Bolsig+} input are collected from Itikawa \cite{Itikawa2005/JPCRD} and Phelps \cite{Phelps/Online} databases.  Particularly, we have used Itikawa cross sections to compute the reaction rates involving water and its derived ions. The electric field $\mathbf{E}$ is computed in the electrostatic approximation according to:
\begin{equation}
	\mathbf{E} = -\nabla\phi,
\end{equation}
and
\begin{equation}
	\label{eq:poisson}
	\nabla^2\phi = -\sum_s\frac{q_s}{\varepsilon_0},
\end{equation}
where $\phi$ is the electrostatic potential, $\varepsilon_0$ the vacuum permittivity and $q_s$ is the electric charge of the species $s$.

The model just described is implemented in Afivo-streamer \cite{Teunissen2017/JPhD},
based on Afivo, an octree-based adaptive mesh refinement framework \cite{Teunissen2018/CoPhC} with OpenMP parallel capabilities and a geometric multigrid solver for Poisson's equation. The \textit{ddr} equations \ref{eq:fluid} are solved using explicit second order time stepping and a flux-limited second order accurate spatial discretization.

\subsection{Initial and boundary conditions}

Fig. \ref{fig:BC_IC} shows a sketch of the axisymmetric domain, the initial and boundary conditions used in the streamer simulations. Our initial conditions consist in a needle-like gaussian electron seed attached to the lower boundary. It has a length $z_0=\SI{5}{\centi\meter}$, a peak value  $n_{e0}=\SI[retain-unity-mantissa = false]{1e21}{\metre^{-3}}$ and an $e$-fold radius $\sigma = \SI{3}{\milli\meter}$. Therefore, the initial electron density reads
\begin{equation}
	n_e =  n_{e,bk} + n_{e0}\exp\left(-\frac{\max\left(z-z_0, 0\right)^2}{\sigma^2} - \frac{r^2}{\sigma^2}\right),
\end{equation}
where $n_{e,bk}$ denotes a uniform background electron density with values $\SI[retain-unity-mantissa = false]{1e9}{\metre^{-3}}$, $\SI[retain-unity-mantissa = false]{1e12}{\metre^{-3}}$ and $\SI[retain-unity-mantissa = false]{1e15}{\metre^{-3}}$ in different simulations. To get a neutral seed we balance the electron density with an equal density of \ce{O_2^+}.  The electron density configuration has been chosen to mimic a needle electrode. Once the simulation starts, the electric field at the tip of the needle-like electron seed is enhanced easing the onset of a streamer. The peak electron density of the seed is high enough to significantly perturb the background electric field but it is also low enough to prevent the time steps in our simulation becoming too short. As for the background electric field, we have applied four evenly-spaced fields in the range \SIrange{7}{10}{\kilo\volt / \centi\meter}. We explore this range to study the behavior of negative and positive streamers around their threshold propagation field, which reflects common conditions both in nature and laboratory experiments. 

Poisson's equation \ref{eq:poisson} is subjected to homogeneous Dirichlet boundary conditions at the top and bottom boundaries and homogeneous Neumann boundary conditions  at the inner and outer boundaries. Homogeneous Neumann boundary conditions are also applied to the equations describing the evolution of the electron and ion densities \ref{eq:fluid}. All the simulations were perform for a temperature of \SI{300}{\kelvin} and a pressure of \SI{1}{atm}.

The minimum grid spacing in our simulations was around \SI{1.5}{\micro \meter}.

\begin{figure}
	\label{fig:BC_IC}
	\centering
	\includegraphics[width=\textwidth]{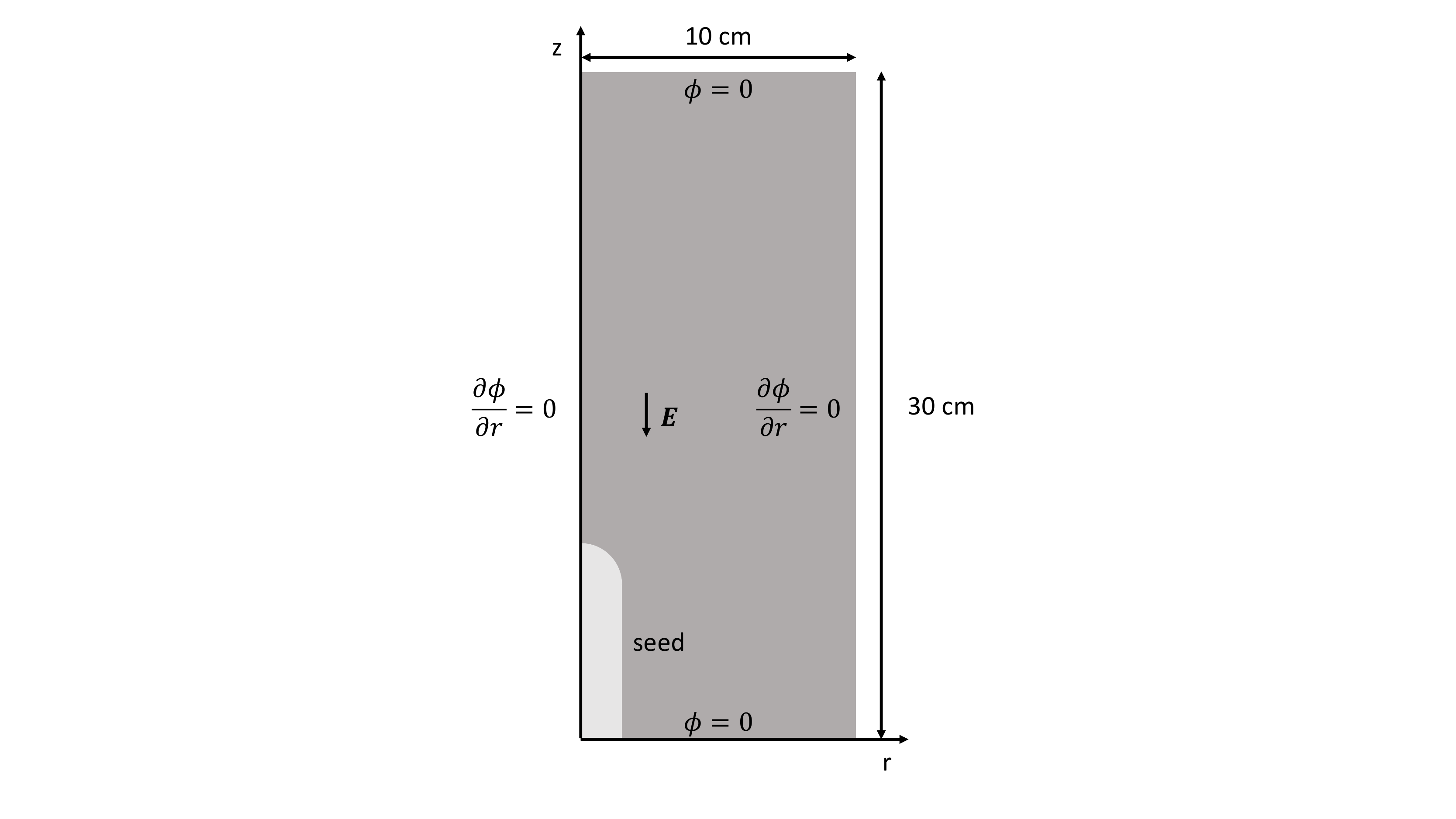}
	\caption{The domain is cylindrically symmetric and is given by the rectangle [0,10]x[0,30] $cm^2$. The electron density has the shape of a needle that extends \SI{5}{\centi\meter} and is placed along the symmetry axis. Poisson's equation \ref{eq:poisson} is solved by applying zero Dirichlet boundary conditions at the top and bottom corner and zero Neumann boundary conditions at the inner and outer boundaries. Note that we also apply a constant background electric field ($\mathbf{E}$) that sets a voltage difference along the domain. The direction of the electric field shown in this figure has only illustration purposes and it will vary in the simulations depending on the polarity of the streamer that we simulate. Homogeneous Neumann boundary conditions are applied to all boundaries for the electron and ion densities \ref{eq:fluid}.}
\end{figure}

\section{Results and discussion}
\label{sec:results}
We simulated the propagation of negative and positive streamers under different absolute humidity \{0, 1.5, 3\} \%, background electron densities $\{10^{9},10^{12},10^{15}\} \unit{m^{-3}}$ and background fields $\{7,8,9,10\} \unit{kV/cm}$. In all cases, streamers propagating in a background density of \SI[retain-unity-mantissa = false]{e9}{\meter^{-3}} are qualitatively similar to those at \SI[retain-unity-mantissa = false]{e12}{\meter^{-3}} and so we focus only on the latter.

\subsection{Positive streamers}
Figure~\ref{fig:ne_evol_pos} shows snapshots at \SIlist{54;71;94}{\nano \second} of positive streamers propagating under different absolute humidity conditions, a background electron density of \SI[retain-unity-mantissa = false]{e12}{\meter^{-3}}, and an electric field of \SI{10}{\kilo \volt / \centi \meter}. From these results, we can infer that the addition of water vapor causes a slower propagation and reduces the radius and conductivity of the streamer channel. These effects are common to all input conditions for our simulations.

\begin{figure}
	\centering
	\includegraphics[width=\textwidth]{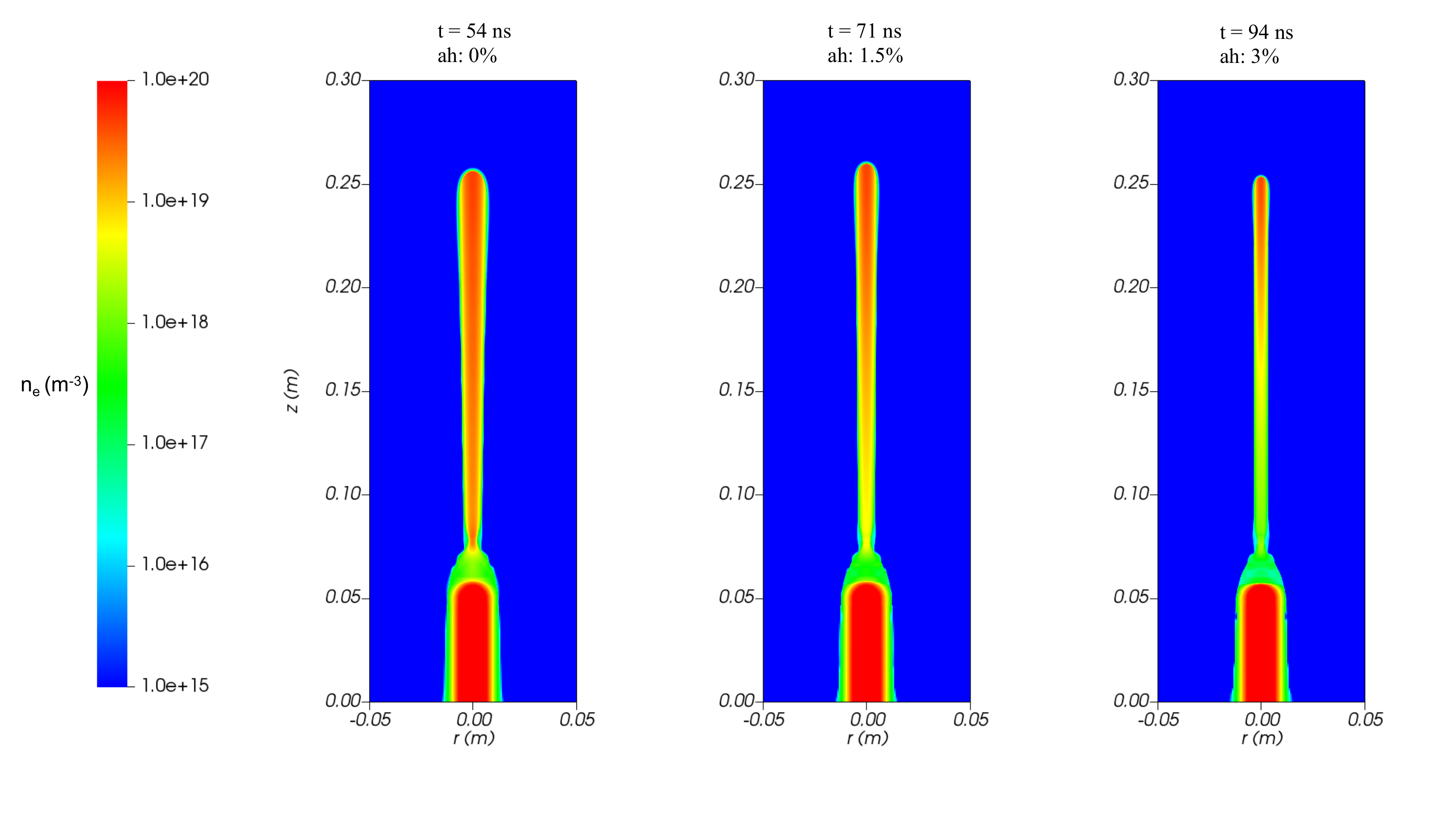}
	\caption{Snapshots of the electron density at \SIlist{54;71;94}{\nano \second} for positive streamers propagating under different absolute humidity conditions (ah ) \{0\%, 1.5\%, 3\%\}, a background electron density of \SI[retain-unity-mantissa = false]{e12}{\meter^{-3}} and a background field of \SI{10}{\kilo \volt / \centi \meter}. Positive streamers in humid air are generally thinner, slower and less conductive.}
	\label{fig:ne_evol_pos}
\end{figure}

\begin{figure}
	\centering
	\includegraphics[width=\textwidth]{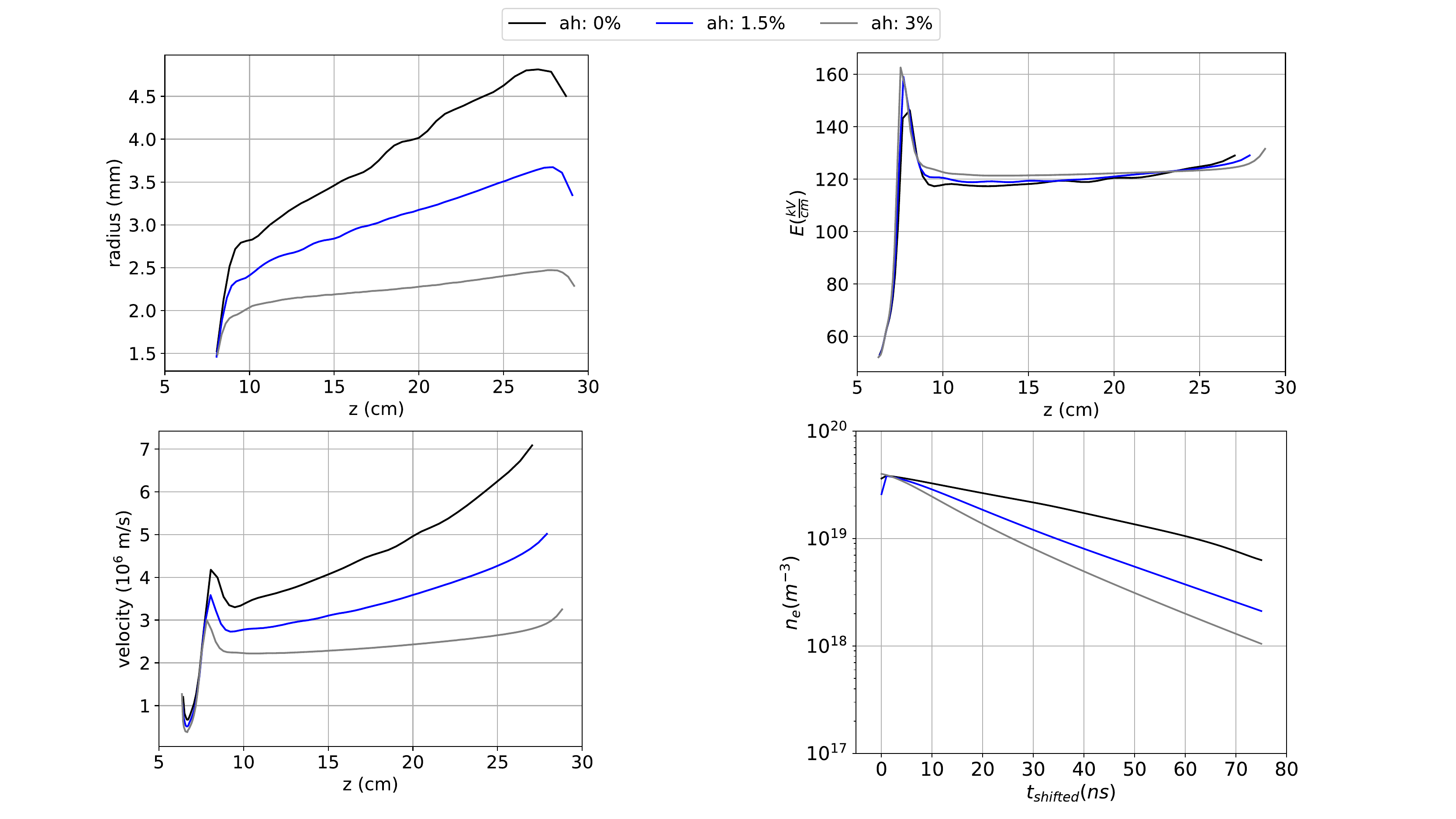}
	\caption{Here we plot the radius of the streamer head, the reduced electric field at the tip, the velocity and the evolution of the electron density at a point of the channel ($r=0, z=\SI{15}{\centi \meter}$) for a positive streamer propagating under different absolute humidity conditions (ah) \{0\%, 1.5\%, 3\%\}, an electron background density of \SI[retain-unity-mantissa = false]{1e12}{m^{-3}} and a background field of \SI{10}{\kilo \volt / \centi \meter}.}
	\label{fig:case1_pos}
\end{figure}

We see a more complete picture of the streamer's evolution in figure~\ref{fig:case1_pos}, where we show the radius and maximum electric field of the streamer head, as well as the propagation velocity and the evolution of the electron density on the streamer channel at a fixed location. We define the streamer head radius as the radius at which the radial component of the electric field attains its maximum and here it increases approximately linearly as the ionized column grows towards the cathode. The streamer velocity also increases with the streamer length but deviates more from a linear trend. On the other hand, the field at the tip remains mostly constant during the propagation. Under saturated water conditions, the radius and velocity at a distance of \SI{25}{cm} from the anode are, respectively, 48\% and 58\% lower than under dry air. 

The addition of water has a significant effect on the evolution of the streamer channel conductivity, which is dominated by electrons. The bottom right panel in figure~\ref{fig:case1_pos} shows the evolution of the electron density at $z = \SI{15}{\centi \meter}$. For easier comparison, the time origin for each line is shifted to match the arrival of the streamer head at the point mentioned. The electron density in the channel \SI{70}{ns} after the streamer passage through this fixed point is nearly six times lower under saturated air than in dry air.  This strong decrease is probably the responsible for the other changes connected with higher humidity. Although the presence of water significantly affects their propagation, with this electric field streamers advance even in saturated humidity conditions.

Another parameter that we explored in our simulations is the background electron density. With an increased background density of \SI{e15}{\meter^{-3}} photoionization becomes less fundamental for the propagation of positive streamers. As before, in figure~\ref{fig:ne_evol_pos_15} we plot different snapshots of the electron density of a positive streamer propagating under this elevated background density and our three values of the  absolute humidity. We see that the effect of adding water vapour is similar to the corresponding case at a lower background density.

\begin{figure}
	\centering
	\includegraphics[width=\textwidth]{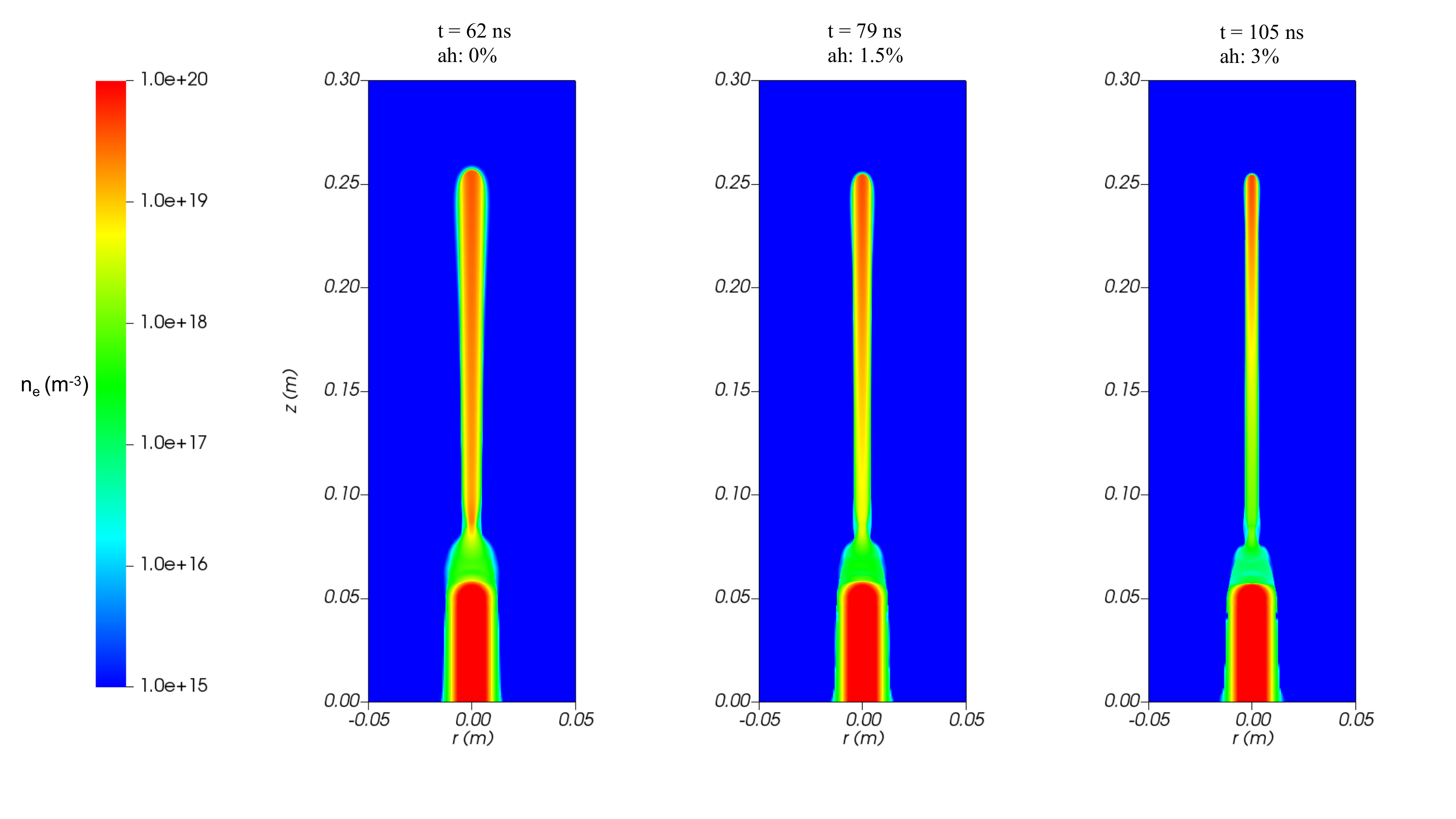}
	\caption{Snapshots of the electron density at \SIlist{62;79;105}{\nano \second} for positive streamers propagating under different absolute humidity conditions (ah) \{0\%, 1.5\%, 3\%\}, a background electron density of \SI[retain-unity-mantissa = false]{e15}{\meter^{-3}} and a background field of \SI{10}{\kilo \volt / \centi \meter}. In humid air, positive streamers are generally thinner, slower and less conductive.}
	\label{fig:ne_evol_pos_15}
\end{figure}

Taking a look in more detail at the radius, maximum electric field and velocity, we can see in figure~\ref{fig:case2_pos} that for a higher background density, all these magnitudes are smaller. Regarding the radius, velocity and channel electron density, variations in humidity lead to deviations of similar magnitude as in the case with lower background density represented in figure~\ref{fig:case1_pos}. The radius and the velocity are lower at most by 48\% and 54\% respectively.  The decrease of electron density at a fixed point is also similar than in figure~\ref{fig:case1_pos}.  However, although, the electric field at the tip is still approximately constant during the streamer propagation, with a higher electron background the electric field value is more sensitive to the humidity and so its value is about \SI{10}{\kilo \volt / \centi \meter} higher in saturated air than in dry air.

\begin{figure}
	\centering
	\includegraphics[width=\textwidth]{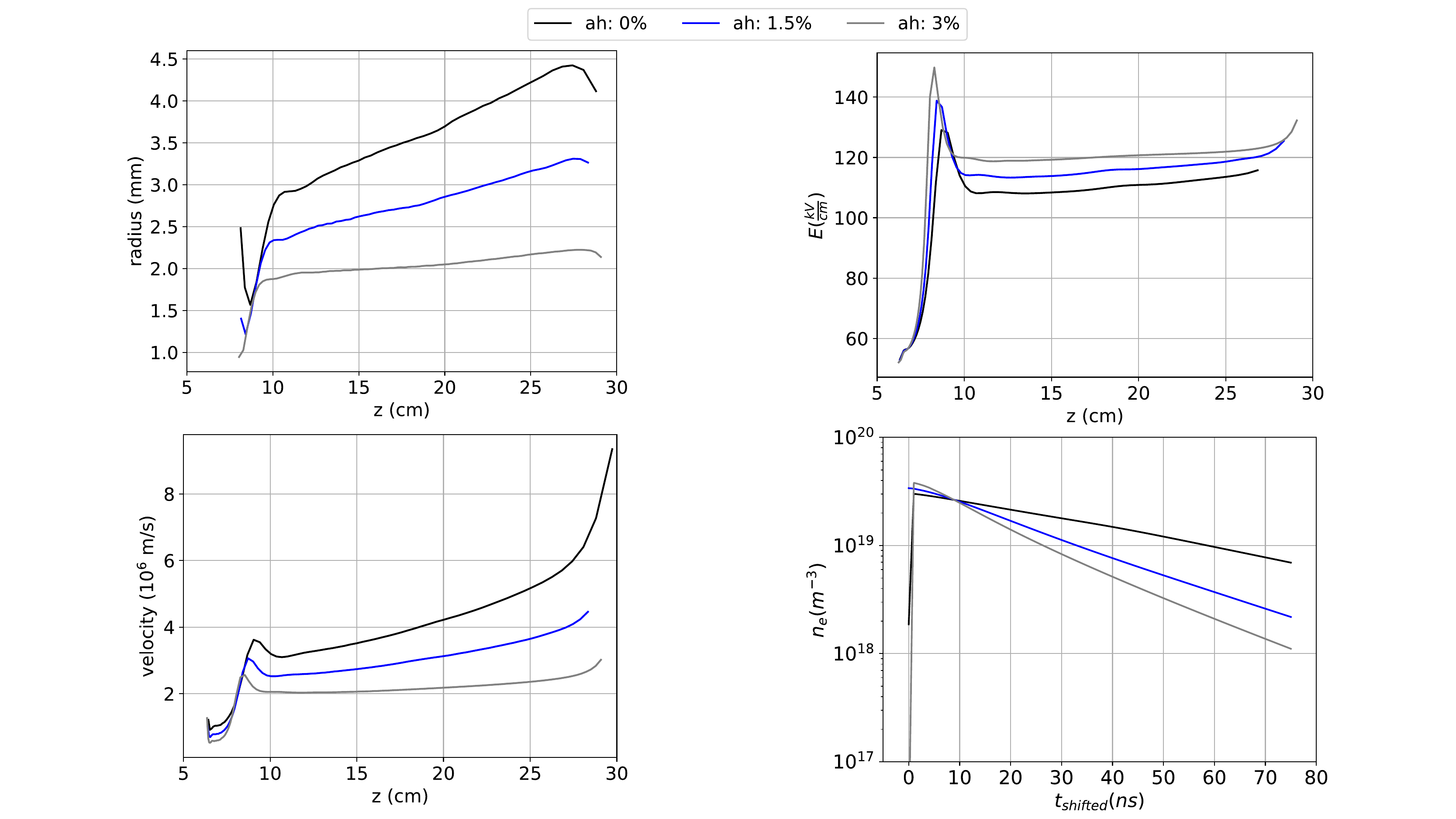}
	\caption{Here we plot the radius of the streamer head, the reduced electric field at the tip, the velocity and the evolution of the electron density at a point of the channel ($r=0, z=\SI{15}{\centi \meter}$) for a positive streamer propagating under different absolute humidity conditions (ah) \{0\%, 1.5\%, 3\%\}, an electron background density of \SI[retain-unity-mantissa = false]{1e15}{m^{-3}} and a background field of \SI{10}{\kilo \volt / \centi \meter}.}
	\label{fig:case2_pos}
\end{figure}

The effect of humidity on positive streamer propagation is even more significant for lower electric fields. As we can see in figure~\ref{fig:ne_evol_pos_15_1}, at \SI{7}{\kilo \volt / \centi \meter} and with a background electron density of \SI{e15}{m^{-3}}, a positive streamer does not propagate under our initial conditions and an absolute humidity of 3\%.  With different initial conditions or with an already existing streamer, it is possible that an electric field of \SI{7}{\kilo \volt / \centi \meter} is sufficient to enable streamer propagation with an humidity of 3\%.  However, since our initial conditions are tailored to facilitate streamer inception, our result suggests that the threshold field for propagation is not much lower than \SI{7}{\kilo \volt / \centi \meter}.  

Propagation is still possible for less humid environments, although it is slower than for dry air and the streamer head radius is noticeably smaller (see figure~\ref{fig:case3_pos}).  At an absolute humidity of 1.5\% the streamer appears close to a steady-state propagation, with roughly constant radius and velocity. We interpret this as an indication that these conditions are close to the streamer propagation threshold.

\begin{figure}
	\centering
	\includegraphics[width=\textwidth]{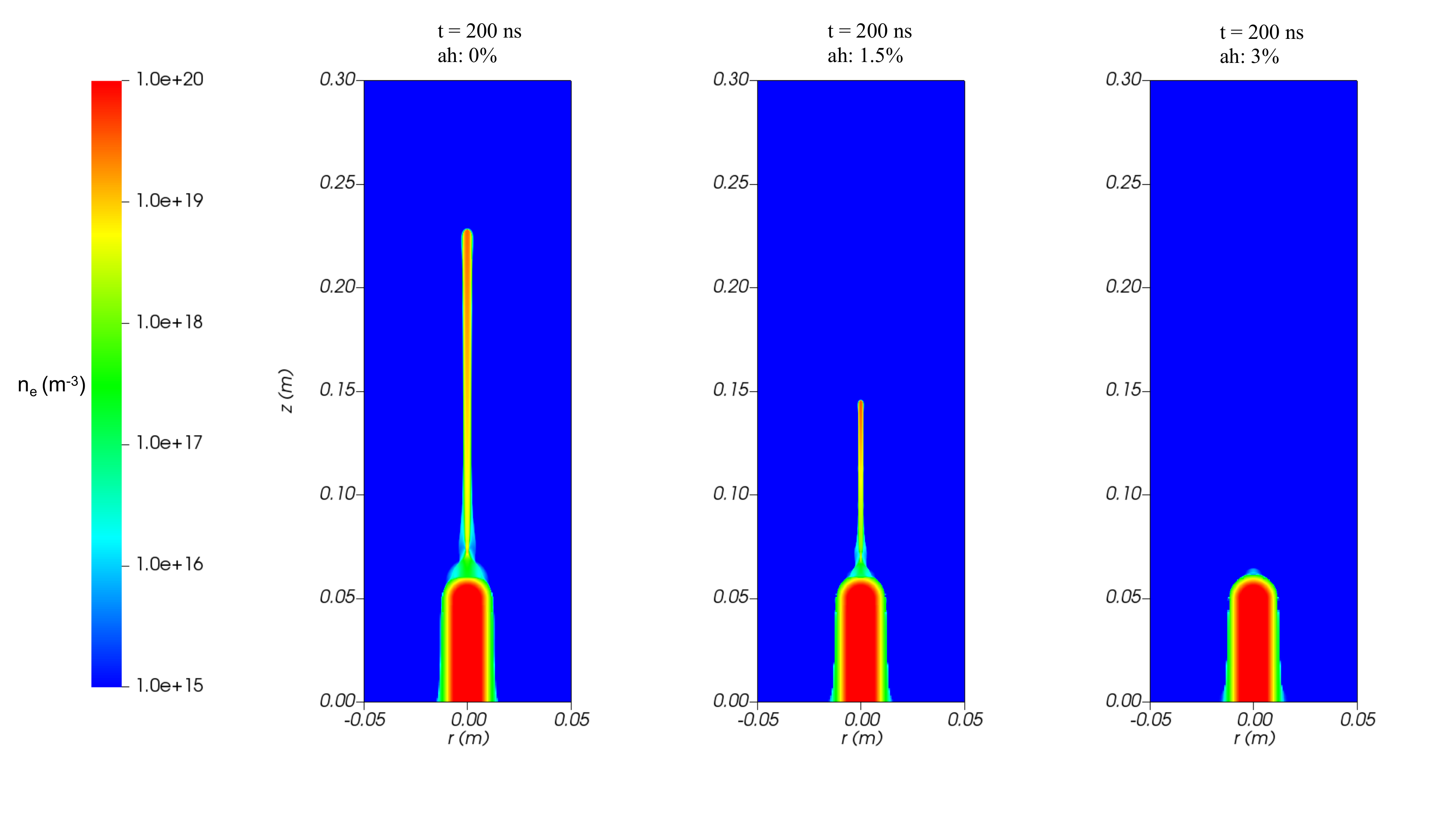}
	\caption{Snapshots of the electron density at \SI{200}{\nano \second} for positive streamers propagating under different absolute humidity conditions (ah) \{0\%, 1.5\%, 3\%\}, a background electron density of \SI[retain-unity-mantissa = false]{e15}{\meter^{-3}} and a background field of \SI{7}{\kilo \volt / \centi \meter}. In saturated air, plot on the right, positive streamer propagation is absent.}
	\label{fig:ne_evol_pos_15_1}
\end{figure}

\begin{figure}
	\centering
	\includegraphics[width=\textwidth]{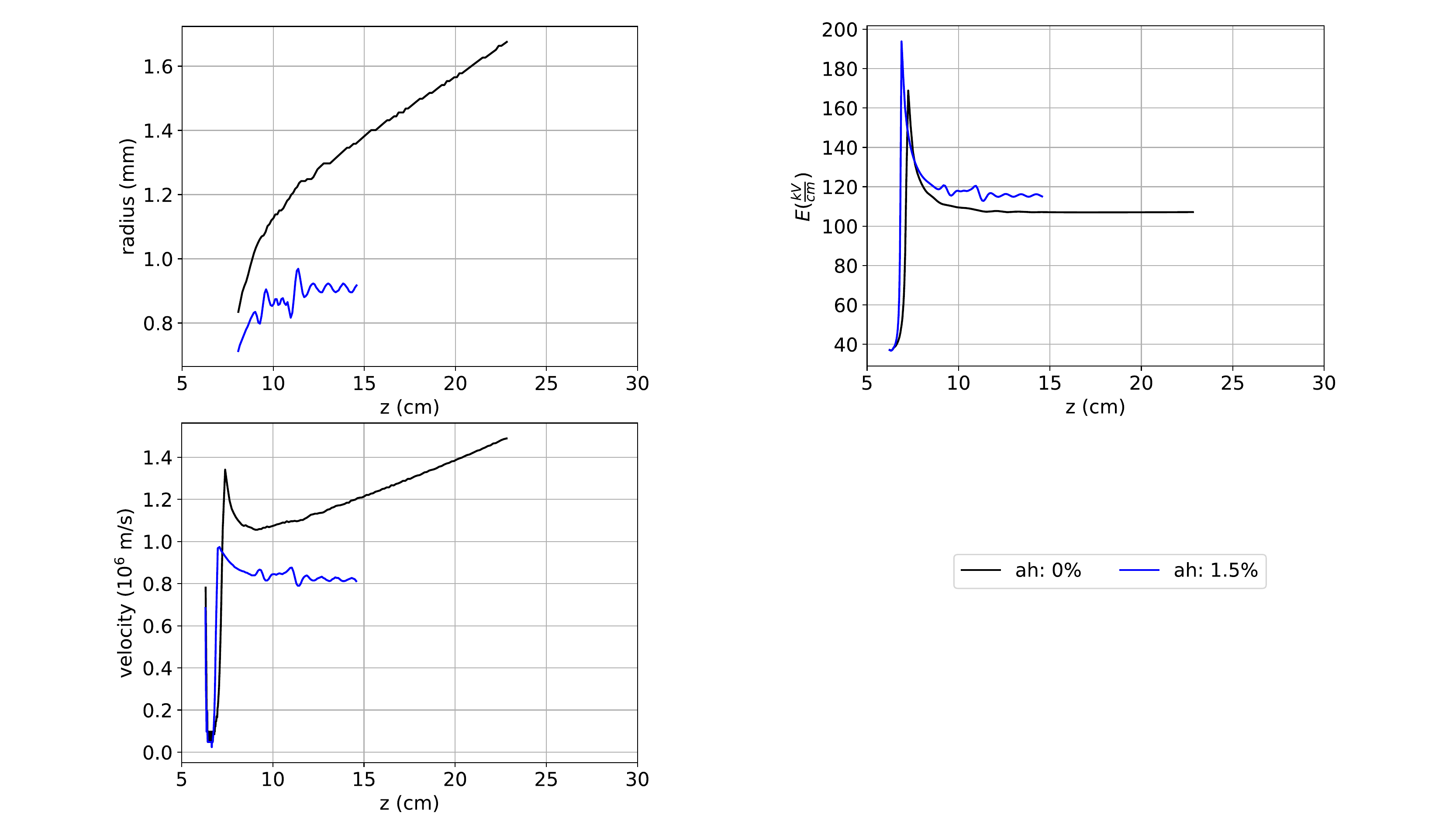}
	\caption{Plots of the radius of the streamer head, the reduced electric field at the tip and the velocity for a positive streamer propagating under different absolute humidity conditions (ah) \{0\%, 1.5\%\}, an electron background density of \SI[retain-unity-mantissa = false]{1e15}{m^{-3}} and a background field of \SI{7}{\kilo \volt / \centi \meter}. Note that we do not plot the results for 3\% humidity because there is no streamer propagation.}
	\label{fig:case3_pos}
\end{figure}

\subsection{Negative streamers}
Let us turn to negative streamers. In general, with this polarity streamers advance only in a more restricted set of conditions, exhibiting a higher threshold field for propagation \cite{Raizer1991/book,Luque2008/JPhD}.

Again, we start our analysis with a background electric field of \SI{10}{kV/cm} and a background electron density of \SI{e12}{m^{-3}}, the results appearing in figures \ref{fig:ne_evol_neg} and~\ref{fig:case1_neg}.  The snapshots of the simulations in figure~\ref{fig:ne_evol_neg} reveal that an increase in humidity leads to decreasing head radius, velocity, and electron density inside the channel. In figure~\ref{fig:case1_neg} we see three regimes for the evolution of the streamer head radius. Whereas in dry air the streamer head expands roughly linearly, with an absolute humidity of 1.5\% the radius stalls at a roughly constant value. Above this value, the negative streamer gets narrower as it propagates towards the anode. We observe a similar behavior in the streamer velocity. On the other hand, both for dry air and a humidity of 1.5\% the streamer maintains the electric field throughout the whole propagation, with a slightly higher value for dry air. Once we surpass an absolute humidity of 1.5\%, we observe that the streamer stagnates in our simulation domain. We attribute this behavior to the fading conductivity inside the streamer channel.

\begin{figure}
	\centering
	\includegraphics[width=\textwidth]{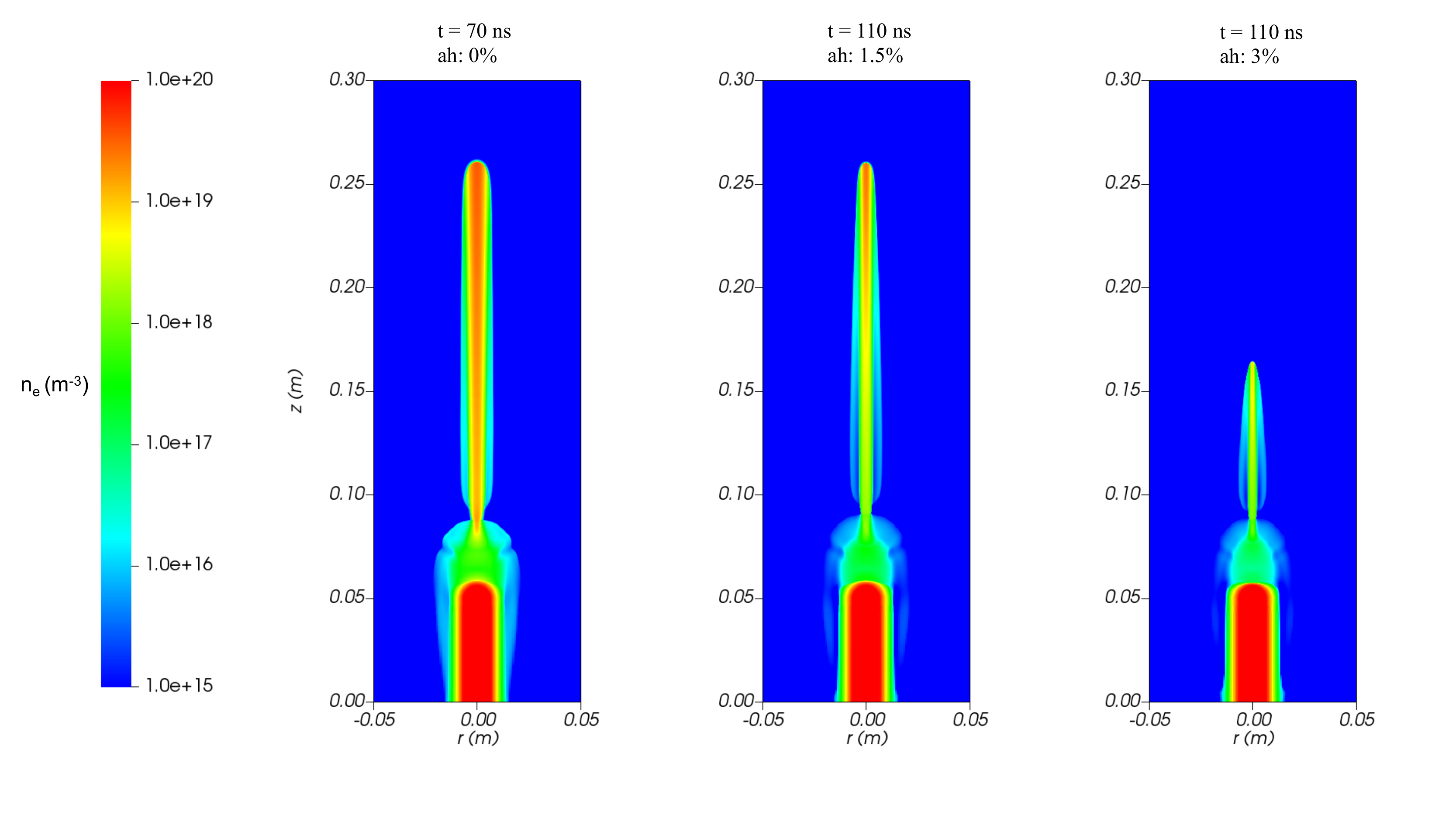}
	\caption{Snapshots of the electron density at \SIlist{70;110;110}{\nano \second} for negative streamers propagating under different absolute humidity conditions (ah) \{0\%, 1.5\%, 3\%\}, a background electron density of \SI[retain-unity-mantissa = false]{e12}{\meter^{-3}} and a background field of \SI{10}{\kilo \volt / \centi \meter}. In humid air, negative streamers are generally thinner, slower and less conductive.}
	\label{fig:ne_evol_neg}
\end{figure}

\begin{figure}
	\centering
	\includegraphics[width=\textwidth]{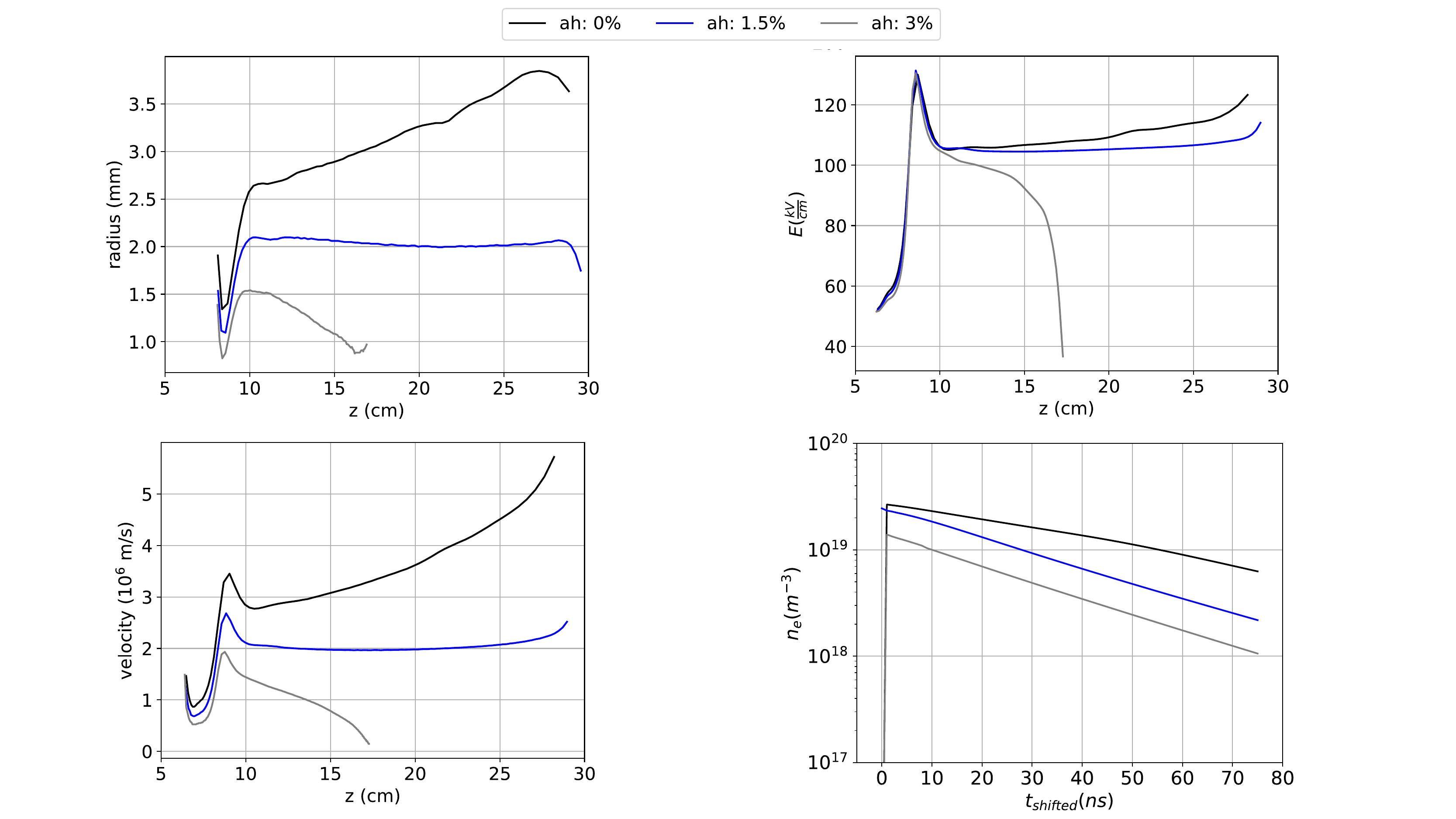}
	\caption{Here we plot the radius of the streamer head, the reduced electric field at the tip, the velocity and the evolution of the electron density at a point of the channel ($r=0, z=\SI{15}{\centi \meter}$) for a negative streamer propagating under different absolute humidity conditions (ah) \{0\%, 1.5\%, 3\%\}, an electron background density of \SI[retain-unity-mantissa = false]{1e12}{m^{-3}} and a background field of \SI{10}{\kilo \volt / \centi \meter}.}
	\label{fig:case1_neg}
\end{figure}

Similar to the positive polarity, a higher background electron density hampers the propagation of a negative streamer. Figures~\ref{fig:ne_evol_neg_15} and~\ref{fig:case2_neg} show results of our simulations with a background of \SI{e15}{m^{-3}} electrons and under \SI{10}{kV/cm}.  Only the case with dry air was capable of bridging the gap.  Again a higher humidity leads to lower radius, velocity and conductivity of the streamer channel. Fig. \ref{fig:case2_neg} shows how in a higher background density, the radius, field and velocity are smaller and the streamers stagnate at shorter time-scales. Note that for dry air and for 1.5\% humidity the conductivity at the streamer channel is similar to the case with a background density of \SI{e12}{m^{-3}} represented in figure~\ref{fig:case1_neg}.

\begin{figure}
	\centering
	\includegraphics[width=\textwidth]{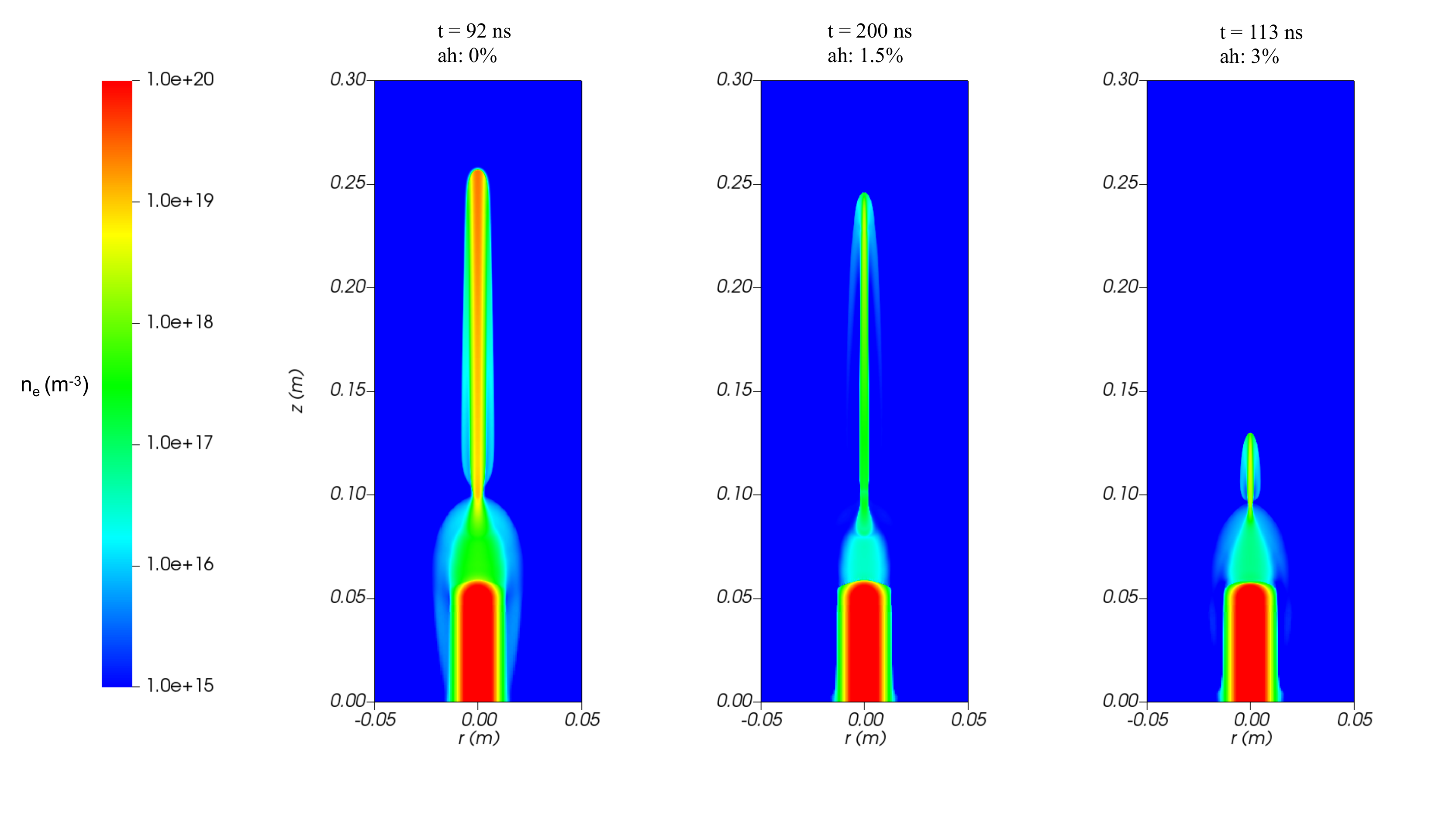}
	\caption{Snapshots of the electron density at \SIlist{92;200;113}{\nano \second} for negative streamers propagating under different absolute humidity conditions (ah) \{0\%, 1.5\%, 3\%\}, a background electron density of \SI[retain-unity-mantissa = false]{e15}{\meter^{-3}} and a background field of \SI{10}{\kilo \volt / \centi \meter}. In humid air, negative streamers are generally thinner, slower and less conductive.}
	\label{fig:ne_evol_neg_15}
\end{figure}

\begin{figure}
	\centering
	\includegraphics[width=\textwidth]{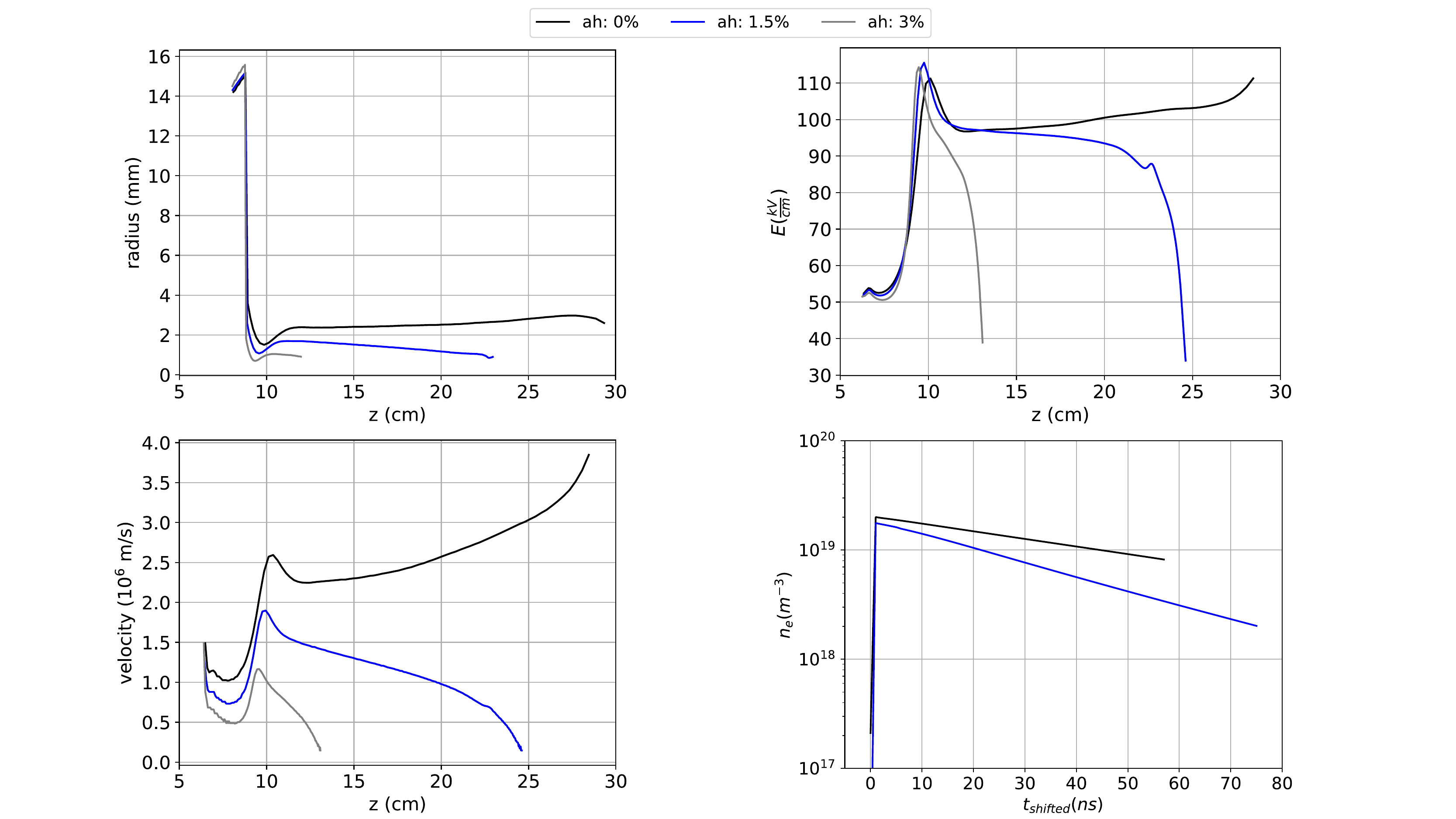}
	\caption{Here we plot the radius of the streamer head, the reduced electric field at the tip, the velocity and the evolution of the electron density at a point of the channel ($r=0, z=\SI{15}{\centi \meter}$) for a negative streamer propagating under different absolute humidity conditions (ah) \{0\%, 1.5\%, 3\%\}, an electron background density of \SI[retain-unity-mantissa = false]{1e15}{m^{-3}}. Note that for an absolute humidity of 3\%, the streamer does not reach the point ($r=0, z=\SI{15}{\centi \meter}$) and therefore, the corresponding evolution of the electron density is not plotted in the right bottom panel.}
	\label{fig:case2_neg}
\end{figure}

\subsection{Discussion}
As we have seen in the simulations described above, the presence of even a small fraction of water molecules strongly affects streamer evolution. Both for positive and negative streamers, the threshold electric field for propagation is higher in humid air, which is consistent with experimental results for positive streamers \cite{Allen1999/JPhD,Allen1989/IEEP}. Furthermore, propagating streamers exhibit markedly lower accelerations and smaller radii. Let us discuss now the origin of these effects.

The presence of water has three major effects on the streamer chemistry:
\begin{enumerate}
	\item Even if the proportion of water is much smaller than that of oxygen, the reaction rate coefficient associated with the three-body attachment (\ref{3body}) where water molecules play the role of the third body is 6-10 times higher \cite{Liu2017/JPhD,Aleksandrov1998/TPhyL}. This implies that for a 3\% humidity, the total attachment rates is a 80\% higher than for dry air under the same electric field.
	\item The recombination rate is enhanced due to the formation of positive water clusters, see figure\ref{fig:pal}.
	\item The formation of negative water clusters prevents electron detachment \cite{Gallimberti1979/JPhys}.
\end{enumerate}
Figure \ref{fig:pal}  confirms the dynamics that we observe, that is, the decay of the electron density in the streamer channel. As we can see, the dominant mechanism is the three-body attachment, followed by the electron recombination with water clusters. Note that the electron detachment from \ce{O2-} is not among the most relevant reactions as this is strongly suppressed by the formation of negative water clusters.

In figure~\ref{fig:nir} we plot the net ionization rate for dry and saturated air as a function of the electric field. We see that in the range of values plotted, the net ionization rate in saturated air is lower than for dry air except in a small region for small electric fields. For energies around \SI{1}{\electronvolt} and below, the excitation of low rotational and vibrational energy levels of water molecules drains electrons energy \cite{Ruiz-Vargas2010/JPhD,Yousfi1996/JAP} and causes the rates of many processes to vanish for low electric fields. Consequently, in humid air, the breakdown electric field is higher and the streamer onset is delayed. In our simulations the electric field inside the streamer channel ranges from \SIrange{5}{10}{\kilo \volt / \centi \meter}. Hence, the evolution of the conductivity is roughly confined to a region of figure~\ref{fig:nir} where the net ionization rate is negative and contains a significant contribution from three-body attachment.

The decaying conductivity due to stronger three-body attachment reduces the amount of charge transported to the streamer head. In extreme cases this charge is insufficient to sustain the streamer propagation. In other cases a propagating solution still exists where the lower charge is compensated by a smaller radius and lower velocity.

\begin{figure}
	\centering
	\includegraphics[width=\textwidth]{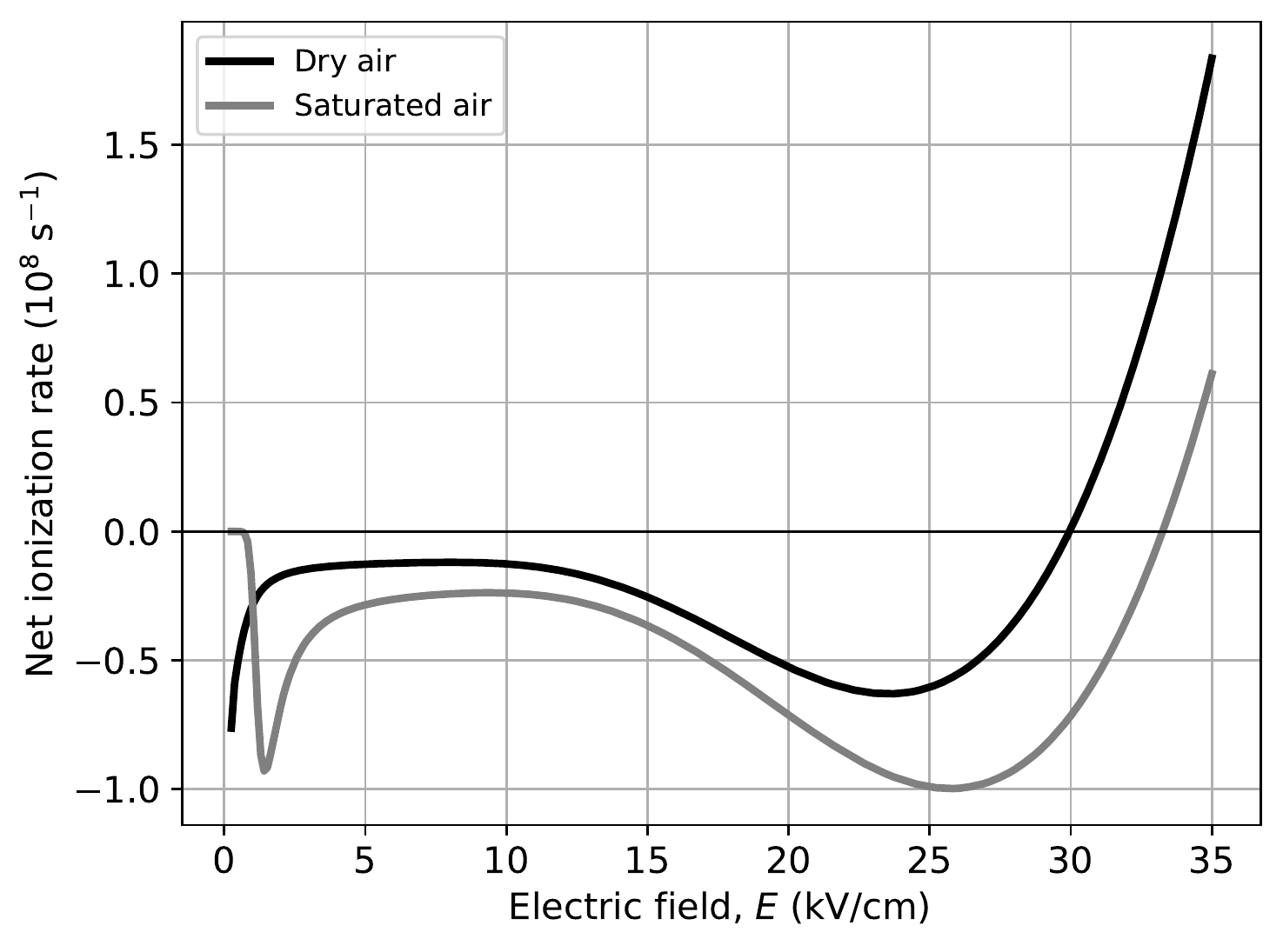}
	\caption{Comparison of the net ionization rate in dry and humid air at atmospheric pressure. The net ionization rate remains smaller in humid air. Hence, the breakdown electric field is higher in humid air, the electron depletion occurs faster and therefore there is a more drastic loss of conductivity that impacts the propagation of the streamer head. This in turn results in a higher electric field threshold for the streamer propagation. The three-body attachment is the main responsible of the electron loss in the electric field range at which the streamer channel is subjected to. This plot is obtained for a temperature of \SI{300}{\kelvin} and a pressure of \SI{1}{atm}.}
	\label{fig:nir}
\end{figure}

\begin{figure}
	\centering
	\includegraphics[width=\textwidth]{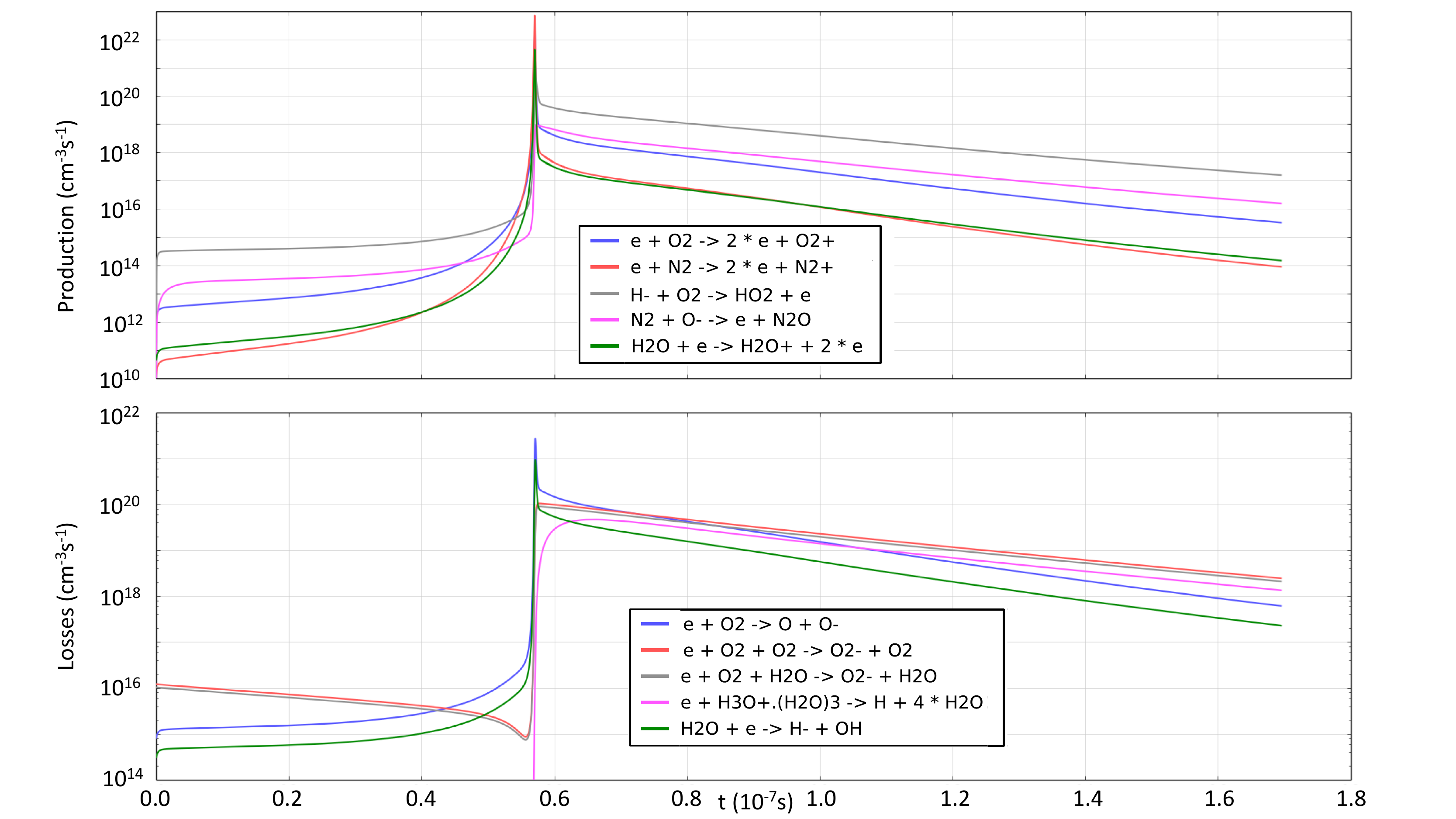}
	\caption{Most relevant electron production and loss mechanisms at (r=\SI{0}{\centi\meter}, z=\SI{15}{\centi\meter}), point located in the channel of a negative streamer propagating in a background electric field of \SI{10}{\kilo \volt/\centi \meter}, a background density \SI[retain-unity-mantissa = false]{1e15}{m^{-3}} and an absolute humidity 1.5\% (see middle plot in Fig. \ref{fig:ne_evol_neg_15}). The streamer channel conductivity is dominated by electrons and so its decay is mostly due to the three-body attachment, followed by the electron recombination with water clusters.}
	\label{fig:pal}
\end{figure}

\section{Conclusions}
\label{sec:conclusions}

We have studied the influence of water vapor content in air on positive and negative streamers under different regimes characterized by diverse background densities and electric fields. These effects are relevant even for a small proportion of water vapor in the air admixture such as 1.5 \% that corresponds to a relative humidity of 50 \%. The most relevant effect of water on the streamer chemistry is the enhancement of the three-body attachment rate. This has two main consequences: on one hand the breakdown electric field increases; on the other, the electron density in the channel depletes faster, restricting more the conditions needed for streamer propagation (i.e. increasing the threshold field for streamer propagation, as stated by reference~\cite{Raizer1991/book} and analyzed experimentally for positive streamers~\cite{Allen1999/JPhD,Allen1989/IEEP}). Under conditions that allow streamers to propagate, higher humidity translates into thinner streamers and a slower propagation. These results agree with other numerical studies of the effect of an enhanced attachment rate in sub-breakdown field regions \cite{Francisco2021/PSST}.

The dynamics of the streamer head and body are linked; adding water makes the streamer propagation more difficult since the net ionization rate is lower in humid air. To compensate this, the streamer tends to reduce the streamer head radius to enhance the focus of the electric field and subsequently increase the ionization at the front. However, the conductivity at the streamer wake quickly reduces in humid air and the streamer channel does not efficiently "transport" the electrode potential to the streamer tip. The streamer head shrinks even further. When the conductivity is low enough and the streamer is unable to focus the electric field at the expenses of reducing its radius, the streamer stagnates as we have seen in our simulations.

In nature, streamer propagation frequently occurs in humid air. Many streamer experiments are also performed in ambient air, with a non-negligible humidity. It is therefore necessary to account for humidity in streamer simulations designed to compare with experiments. The effects that we have mentioned on the streamer propagation also have an imprint on larger scale discharges. Because the presence of water can lead to streamer stagnation, the dimensions of a streamer corona may be affected too. Besides, in humid air, the streamer channel decays on shorter time-scales. This probably affects the rate at which subsequent streamer coronas launch at the leader tip in long spark discharges. Finally, in the streamer propagation the interplay between streamer head and streamer wake leads to a dynamic streamer head radius. The evolution of the radius is related to streamer branching \cite{Briels2008/JPhD} and so humidity is a factor to be considered when we estimate the number of streamer branches per unit length in a streamer corona.

\section{Data availability statement}
The data that support the findings of this study are openly available at the following URL/DOI: \url{https://doi.org/10.5281/zenodo.6699969}.

\section{Acknowledgments}
This work was supported by Junta de Andalucía under the project PY20-00831. A. Malagón-Romero acknowledges a postdoctoral contract under the project PY20-00831. A. Malagón Romero and A. Luque acknowledge the State Agency for Research of the Spanish MCIU through the ``Center of Excellence Severo Ochoa'' Award for the Instituto de Astrofísica de Andalucía (SEV-2017-0709). The simulation code for this work is available at \url{https://gitlab.com/MD-CWI-NL/afivo-streamer}. The version used for the simulations corresponds to the commit 1ff2676ba48a5eb568f06c7b11a548629a5ff20c. All the simulations were carried out in the MareNostrum Supercomputer under project number FI-2021-2-0025, granted by the Barcelona Supercomputing Center.

\newcommand{\nat}{Nature}
\newcommand{\pre}{Phys. Rev. E}
\newcommand{\prb}{Phys. Rev. B}
\newcommand{\physrep}{Phys. Rep.}
\providecommand{\newblock}{}

\end{document}